\documentstyle[pra,epsf,aps,twocolumn]{revtex}
\def\lfrac#1#2{#1/#2}
\def\sbf#1{{\footnotesize {\bf #1}}}

\begin{document}
\sloppy
\title{Calculation of Second Topological Moment $\langle m^2 \rangle$
of Two Entangled Polymers}
%
%
\author{Franco Ferrari$^{(1)}$\thanks{fferrari@science.unitn.it}
Hagen Kleinert$^{(2)}$\thanks{kleinert@physik.fu-berlin.de}
and Ignazio Lazzizzera$^{(1)}$\thanks{lazi@tn.infn.it}\\
{$^{(1)}$\it Dipartimento di Fisica, Universit\`a di Trento, I-38050 Povo,
Italy\\
and INFN, Gruppo Collegato di Trento.}\\
{$^{(2)}$\it Institut f\"ur Theoretische Physik,\\
Freie Universit\"at Berlin, Arnimallee 14, D-14195 Berlin, Germany.}}
\date{March 1999}
\maketitle
\begin{abstract}
We set up a Chern-Simons theory for the entanglement
of two polymers $P_1$ and $P_2$, and
calculate
the second topological moment $\langle m^2 \rangle$,
 where $m$ is the linking number.
The result approximately to a polymer
in an ensemble of many others, which are considered as
a single very long effective polymer.
\end{abstract}
\section{The Problem}
\begin{figure}[tbhp]
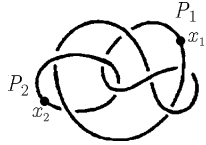

\input link.tps
\caption[]{Closed polymers $P_1, P_2$ with trajectories $C_1, C_2$
respectively.}
\label{Fig. 1}\end{figure}

Consider two polymers  $P_1$ and $P_2$
which statistically can be linked with each other any  number of times $m=0,1,2,\dots~$.
The situation is illustrated in Fig.~\ref{Fig. 1}
for $m= 2$.
We would like to find the probability
distribution of the linking numbers $m$ as a function of
the lengths of $P_1$ and $P_2$.
As a first contribution to solving this problem
we calculate,
in
this note,  the second moment of this distribution,
$\langle m^2 \rangle$. The self-entanglements
are ignored.

The solution of this two-polymer problem
may be considered as an approximation
to a more interesting physical problem
in which
a particular polymer
is
linked to
any number
 $N$ of polymers, which are effectively replaced by a single long
``effective''
polymer \cite{BS}.

Let
$G_m ({\bf x}_1, {\bf x}_2; L_1, L_2)$ be
the configurational probability to
find the polymer $P_1$ of length $L_1$
with fixed coinciding end points at ${ \bf x}_1$
and
the polymer $P_2$ of length $L_2$
with fixed coinciding end points at ${ \bf x}_2$,
entangled with
a Gaussian linking number $ m$.

The second moment
 $\langle m^2 \rangle$ is defined by the ratio of integrals
\begin{eqnarray}
  \langle m^2 \rangle = \frac{\int d^3 {\bf x}_1 d^3 {\bf x}_2
     \int^{+\infty}_{-\infty } dm ~m^2 G_m
\left({\bf x}_1, {\bf x}_2; L_1, L_2\right)}
 {\int d^3 {\bf x}_1d^3 {\bf x}_2 \int^{+\infty}_{-\infty} dm\, G_m
    \left( {\bf x}_1, {\bf x}_2; L_1, L_2\right) }
\label{1}\end{eqnarray}
performed for either of the two probabilities.
The integrations in $d^3  {\bf x}_1d^3 {\bf x}_2 $ covers all positions
of the end points.
The denominator plays the role of a partition function
of the system:
\begin{eqnarray}
Z\equiv
 {\int d^3 {\bf x}_1 d^3 {\bf x}_2 \int^{+\infty}_{-\infty} dm\, G_m
    \left( {\bf x}_1, {\bf x}_2; L_1, L_2\right) }
\label{1Z}\end{eqnarray}

Due to the translational invariance of the system, the probabilities
depend only on the differences between the end point coordinates:
\begin{eqnarray}
 G_m \left( {\bf x}_1, {\bf x}_2; L_1, L_2\right) = G_m \left({\bf x}_1
 - {\bf x}_2;
L_1, L_2\right)
\label{2}\end{eqnarray}
Thus, after the shift of variables,
the spatial double integrals in (\ref{1}) can be
rewritten as
\begin{eqnarray}
&&\!\!\!\!\!\!\!\!\!\!\!\!\!\!\int d^3 {\bf x}_1 d^3 {\bf x}_2 G_m ({\bf x}_1
  - {\bf x}_2 ; L_1, L_2)
\nonumber \\&& =
V\int d^3 {\bf x}  G_m ({\bf x} ; L_1, L_2),
\label{3}\end{eqnarray}
where $V$ denotes the total volume of the system.

\section{Polymer Field Theory for Probabilities}

The linking number for the two polymers
is given by the Gauss integral
\begin{eqnarray} \label{gauss}
I_G(P_1,P_2) = \frac{1 }{4\pi }
 \oint_{P_1} \oint_{P_2}
  [d{\bf x}_1 \times  d{\bf x}_2]
\cdot \frac{{\bf x}_1-{\bf x}_2}{ \vert {\bf x}_1 - {\bf x}_2\vert ^3}.
\end{eqnarray}
It takes the values $m=0,\,\pm1,\,\pm2,\,\dots~$.
With the help of two vector potentials
 ${\bf A}_1$
and ${\bf A}_2$, the phase factor $e^{im \lambda }$
 can be obtained as a result of a local
gauge theory of the Chern-Simons type:
\begin{eqnarray}
e^{im \lambda }&=&
\int {\cal D}A_1^\mu
 {\cal D}A_2^\mu\nonumber \\
&\times&  e^{-{\cal A}_{\rm CS}
- \kappa \int_{P_1} d{\sbf{x}}_1\cdot {\sbf{A}}_1
- \lambda \int_{P_2} d{\sbf{x}}_2\cdot {\sbf{A}}_2},
\label{@imlam}
\end{eqnarray}
where ${\cal A}_{\rm CS}$ is the action
\begin{eqnarray}
 {\cal A}_{\rm CS}
 =i
{ \kappa }
\int d^3 {\bf x} \,\varepsilon_{\mu \nu  \rho }
   A_1^\mu \partial _ \nu  A_2^\rho           ,
\label{5CS}\end{eqnarray}
Indeed, the correlation functions $D^{\mu \nu }_{ij}({\bf x},{\bf x}')$ of the gauge fields are
\begin{eqnarray}
\langle A_1^\mu({\bf x}) A_1^ \nu ({\bf x}')\rangle&\!\!=\!\!&0,~~
\langle A_2^\mu({\bf x}) A_2^ \nu ({\bf x}')\rangle=0,~\\
\langle A_1^\mu({\bf x}) A_2^ \nu ({\bf x}')\rangle&=&
\int \frac{d^3p}{(2\pi )^3}
e^{i{\sbf p}({\sbf x}-{\sbf x}')}
\frac{ i\epsilon _{\mu   \lambda \nu  } {k^ \lambda } }{{\bf k}^2}
\label{@7} \\
&=&\frac{1}{4\pi }\epsilon _{\mu \lambda  \nu }\nabla _ \lambda \frac{1}{|{\bf x}-{\bf x}'|}
\nonumber \\
&=&\frac{1}{4\pi }\epsilon_ {\mu \nu  \kappa }\frac{({ x}-{x}')^ \lambda }{|{\bf x}-{\bf x}'|^3},
\label{@9}\end{eqnarray}
such that the functional integral
on the right-hand side of (\ref{@imlam})
produces directly the phase factor $e^{iI_G(P_1,P_2) \lambda }$
with the eigenvalue $e^{im \lambda }$.

We can select configurations with a certain linking number
$m$ from all configurations by
forming the integral $
\int^{\infty}_{-\infty}  d  \lambda  e^{-im \lambda }$ over this quantity.

The most efficient way of describing the statistical
fluctuations of the
polymers $P_1$ and $P_2$ is by
two complex polymer fields
$ \psi_1^{{ a}_1} ({\bf x}_1)$ and
 $\psi_2^{{ a}_2} ({\bf x}_2)$
with $n_1$ and $n_2$ replica $(a_1=1,\dots,n_1,~a_2=1,\dots,n_2)$.
At the end we shall take $n_1,n_2\rightarrow 0$
to ensure that these  fields describe only one polymer each \cite{pi}.
For these fields we define an auxiliary
probability $G_ \lambda  (\vec {\bf x}_1, \vec{\bf x}_2 ;  \vec{z} )$
to find the polymer $P_1$
with open ends  at ${\bf x}_1,{\bf x}_1'$
and the polymer $P_2$
with open ends  at ${\bf x}_2,{\bf x}_2'$.
The double vectors
$\vec{{\bf x}}_1\equiv ({\bf x}_1,{\bf x}_1')$ and
$\vec{{\bf x}}_2\equiv ({\bf x}_2,{\bf x}'_2)$
collect initial and final
endpoints of the two polymers
  $P_1$ and $P_2$.
The
auxiliary
probability $G_ \lambda  (\vec {\bf x}_1, \vec{\bf x}_2 ;  \vec{z} )$
is given
by a functional integral
\begin{eqnarray}
&&\!\!\!\!\!\!\!\!\!\!\!G_ \lambda  (\vec{{\bf x}}_1, \vec{{\bf x}}_2 ; \vec z )
  =   \lim _{n_1, n_2 \rightarrow 0} \int
  {\cal D} ({\rm fields}) \,\nonumber \\
 & & \times \psi_1^{{a}_i} ({\bf x}_1)
  \psi_1^{* { a}_1}  ({\bf x}_1')
\psi_2^{{ a}_2} ({\bf x}_2) \psi_2^{*{ a}_2} ({\bf x}_2')
    e^{- {\cal A}}
\label{4}\end{eqnarray}
where
${\cal   D} (\mbox{fields})$
indicates the measure of functional
integration, and
 ${\cal A}$ the action
governing the fluctuations.
It consists of kinetic terms for the fields,
a quartic interaction of the fields to account
for the fact that two monomers of the polymers
cannot occupy the same point, the so-called {\em excluded-volume effect\/},
and a Chern-Simons field to
describe the linking number $m$.
Neglecting at first
the   excluded-volume effect and focusing attention
on the linking problem  only, the action reads
\begin{equation}
{\cal A}={\cal A}_{\rm CS}+{\cal A}_{\rm pol},
\label{@5}\end{equation}
with a polymer field action
\begin{eqnarray}
 {\cal A}_{\rm pol} =  \sum _{i=1}^{2} \int d^3{\bf x} \left[ |\bar {\bf D}^i \Psi_i|^2 +
 m^2_i |\Psi_i |^2 \right].
\label{5}\end{eqnarray}
 in which we have omitted a
gauge fixing term, which enforces the  Lorentz gauge.
They are coupled to the polymer fields by the
 covariant derivatives
\def\nablab{\BF \nabla}
\newcommand{\BF}[1]{\mbox{\boldmath $#1$}}
\begin{eqnarray}
 {\bf D}^i = {\nablab} +i  \gamma _i {\bf A}^i,
\label{7}\end{eqnarray}
with the coupling constants $ \gamma _{1,2}$
given by
\begin{eqnarray}
  \gamma _1 =  \kappa  ~~~~~~
   \gamma _2 =  \lambda .
\label{8}\end{eqnarray}
The square masses of the polymer fields contain the chemical
potentials $z_{1,2}$ of the polymers:
\begin{eqnarray}
   m_i^2 = 2 M z_i.
\label{9}\end{eqnarray}
 They are  conjugate variables to the length
parameters $L_1$ and $L_2$, respectively.
The symbols $\Psi_i$ collect the replica of the
two polymer fields
 \begin{eqnarray}
 \Psi_i = \left( \psi^1_i , \dots , \psi_i^{n_i} \right), ~~~\Psi_i^*
=
     \left( \psi_i^{* 1 }, \dots , \psi_i^{* n_i}\right),
\label{10}\end{eqnarray}
and their absolute squares
contain the sums over the replica
\begin{eqnarray}
 |   {\bf D}^i \bar \Psi_i |^2 = \sum_{a_i =1}^{n_i}
    | {\bf  D}^i  \psi_i^{a_i} |^2 ,
 ~~~ |\Psi_i |^2 = \sum_{a_i=1}^{n_i}
       | \psi_i^{a_i}|^2.
\label{11}\end{eqnarray}
Having specified the fields,
we can now write down the measure of functional integration
in Eq.~(\ref{4}):
\begin{eqnarray}
  {\cal D} (\mbox{fields}) = \int  {\cal D}  A_1^\mu {\cal D} A_2^\mu
	    {\cal D} \Psi_1 {\cal D}\Psi_1^* {\cal D} \Psi_2 {\cal D} \Psi_2^* .
\label{12}\end{eqnarray}
 By Eq.~(\ref{@imlam}), the parameter $ \lambda $ is conjugate to the linking number $m$.
We can therefore calculate the desired probability
  $G_m (\vec {\bf x}_1, \vec {\bf x}_2; L_1, L_2)$
from the auxiliary one
  $G_ \lambda  (\vec {\bf x}_1, \vec {\bf x}_2; \vec z)$
by the following
 Laplace integrals
$\vec  z=(z_1,z_1)$:
\begin{eqnarray}
&&G_m (\vec {\bf x}_1, \vec {\bf x}_2; L_1, L_2)
 = \lim_{{\bf x}_1' \rightarrow {\bf x}_1 \atop
              {\bf x}_2' \rightarrow {\bf x}_2}
 \int^{c+ i\infty}_{c - i \infty} \frac{dz_1}{2\pi i}\frac{dz_2}{2\pi i}
   e^{(z_1 L_1 + z_2 L_2)} \nonumber \\
    &&~~~~~~~~~~~~\times
\int^{\infty}_{-\infty}  d  \lambda  e^{-im \lambda }
	  G_ \lambda  \left( \vec{\bf x}_1 ,\vec{\bf x}_2 ; \vec z\right).
\label{13}\end{eqnarray}
%
\section{Calculating the Partition Function}
Let us use the polymer field theory
to calculate the partition function (\ref{1Z}).
 By Eq.~(\ref{13}), it is given by the integral
over the auxiliary probabilities
\begin{eqnarray}
 Z & = &\int d^3 {\bf x}_1 d^3 {\bf x}_2 \lim_{{\bf x}_1' \rightarrow {\bf x}_1
 \atop
   {\bf x}_2 '\rightarrow  {\bf x}_2}
   \int^{c + \infty}_{c - i \infty} \frac{d z_1}{2\pi i}
   \frac{dz_2}{2\pi i} e^{(z_1 L_1 + z_2 L_2)} \nonumber \\
&&\times \int^{+ \infty}_{-\infty} d m \int^{+ \infty}_{-\infty} d \lambda e^{-im \lambda }
     G_ \lambda \left( \vec{\bf x}_1 ,\vec{\bf x}_2 ; \vec z\right).
\label{15}\end{eqnarray}
  The integration over $dm$ is trivial and gives
 $2  \pi   \delta ( \lambda )$, enforcing $ \lambda =0$, so that
\begin{eqnarray}
 Z & = &- \int d^3 {\bf x}_1 d^3 {\bf x}_2 \lim_{{\bf x}_1 \rightarrow {\bf x}_1
 \atop {\bf x}_2' \rightarrow {\bf x}_2}  \int^{c + i\infty}_{c - i \infty}
     \frac{dz_1 dz_2}{2\pi i} e^{(z_1 L_1 + z_2 L_2)} \nonumber \\
  & &~~~~~~~~~~~~~\times~~
    G_{ \lambda =0} \left( \vec{\bf x}_1 ,\vec{\bf x}_2 ; \vec z    \right)
\label{16}\end{eqnarray}
To compute $
 G_{ \lambda =0} \left( \vec{\bf x}_1 ,\vec{\bf x}_2 ; \vec z
    \right)$ we observe
 that the action ${\cal A}$
 in Eq.~(\ref{@5})  is quadratic
in $ \lambda $. Let us expand ${\cal A}$ as
\begin{eqnarray}
  {\cal A} = {\cal A}_0 +  \lambda  {\cal A}_1 +  \lambda ^2 {\cal A}_2
\label{17}\end{eqnarray}
where
\begin{eqnarray}
{\cal A}_0 &\equiv  &{\cal A}_{\rm CS}  \nonumber \\
&+& \int d^3 {\bf x} \left[ |{\bf D}_1 \Psi_1|^2 + | \nablab \Psi_2
  |^2 + \sum_{i =1} ^2  |\Psi_i|^2 \right],
\label{18}
\end{eqnarray}
a linear coefficient
\begin{eqnarray}
 {\cal A}_1  \equiv \int d^3 {\bf x} \,\,{\bf j}_2{ ({\bf x}}) \cdot
   {\bf A}_{2} ({\bf x})
\label{19}\end{eqnarray}
 with a
``current" of the second polymer field
\begin{eqnarray}
{\bf j}_2 ({\bf x}) = i \Psi_2^*({\bf x} ){ \nablab} \Psi_2({\bf x} ),
\label{20}\end{eqnarray}
and a quadratic coefficient

\begin{eqnarray}
 {\cal A}_2 \equiv \frac{1}{4} \int {d^3{\bf x}}~
    {\bf A}_2^2 |\Psi_2 ({\bf x})|^2 .
\label{21}\end{eqnarray}
With these definitions
we can rewrite
(\ref{17}) as
\begin{eqnarray}
 &&\!\!\!\!\!\!\!\!\!\!\!\!\!G_{ \lambda =0} \left( \vec{\bf x}_1 ,
\vec{\bf x}_2 ; \vec z
    \right)
   =  \int {\cal D} ({\rm fields}) e^{-{\cal A}_0} \nonumber \\
&& \times  \psi_1^{{a}_1}({\bf x}_1)
      \psi_1^{*{a}_1} ({\bf x}_1')\psi_2^{{a}_2} ({\bf x}_2) \psi_2^{{a}_2} ({\bf x}')
\label{22}\end{eqnarray}
From Eq.~(\ref{18}) it is clear that
$G_{ \lambda =0} \left( \vec{\bf x}_1 ,\vec{\bf x}_2 ; \vec z
    \right)$
is the product of the configurational
probabilites of two free polymers.

Note that the fields $\Psi_2, \Psi_2^*$ are free, whereas
the fields $\Psi_1, \Psi^*_1$ are apparently not free because
of the couplings with the Chern-Simons fields through the covariant
derivative ${\bf  D}^1$.
This is, however, an illusion:
the fields $A^i_\mu$ have a vanishing
diagonal propagators $ \langle A^i_\mu A^i_ \nu \rangle =0 $.
 integrating out $A_2^\mu$ in (\ref{22}), we find
the flatness condition:
\begin{equation}
 \varepsilon^{\mu \nu  \rho } \partial_ \nu  A^i_\mu= 0.
\label{23}\end{equation}
 On a flat space with vanishing boundary conditions at infinity this
implies $ A_1^\mu  = 0$.
As a consequence,
the functional
integral (\ref{22}) factorizes
\begin{eqnarray}
\!\!\!
   G_{ \lambda =0} \left( \vec{\bf x}_1 ,\vec{\bf x}_2 ; \vec z    \right)
 = G_0 ({\bf x}_1 - {\bf x}_1'; z_1) G_0 ({\bf x}_2  - {\bf x}_2'
    ; z_2 )   ,
\label{24}\end{eqnarray}
where $ G_0 ({\bf x}_i - {\bf x}_i' ; z_i)$
are the free correlation functions of the polymer fields:
\begin{eqnarray}
 G_0 ({\bf x}_i - {\bf x}_i' ; z_i) =
 \langle \psi_i^{{a}_i} ({\bf x}_i ) \psi_i^{*{a}_i}
    ({\bf x}_i') \rangle.
\label{25}\end{eqnarray}
  In
 momentum
space, the correlation functions are
\begin{eqnarray}
 \langle \tilde \psi^{{a}_i} ({\bf k}_i)
\tilde\psi_i^{*{a}_i}  ({\bf k}_i')\rangle =
     \delta^{(3)}  \left({\bf k}_i + {\bf k}_i' \right)
    \frac{1}{{\bf k}_i^2 + m_i^2}
\label{26}\end{eqnarray}
such that
\begin{eqnarray}
 G_0 ({\bf x}_i - {\bf x}_i' ; z_i) =
 \int \frac{d^3 { k}}{(2\pi)^3}
  e^{i{\bf k} \cdot {\bf x}}
    \frac{1}{{\bf k}_i^2 + m_i^2},
\label{@}\end{eqnarray}
				  and
\begin{eqnarray}
  G_0 ({\bf x}_i - {\bf x}_i'; L_i )
&=&\int^{c+i\infty}_{c-i\infty} \frac{dz_i}{2\pi i}e^{z_iL_i} G_0
   ({\bf x}_i - {\bf x}_i' ; z_i )
   \nonumber \\
 &  &\!\!\!\!\!\!\!\!\!\!\!\!\!\!\!\!\!\!\!\!\!\!\!\!\!\!\!\!\!\!\!\!\!
= \frac{1}{4  \sqrt{2}M } \left(\frac{M}{2 \pi }\right)^{3/2}
   L_i^{-3/2} e^{{- M({\bf x} _i -{\bf x}_i')}/{2L_i} }.
\label{27}\end{eqnarray}
Thus we obtain for  (\ref{16}):
\begin{eqnarray}
\lefteqn{
\!\!\!\!\!\!\!\!\!\!\!\!\!\! Z = 2  \pi  \int d^3 {\bf x}_1 d^3 {\bf x}_2  }\nonumber \\
   && \times   \lim_{{\bf x}_1' \rightarrow {\bf x}_1 \atop
                 {\bf x}_2'\rightarrow {\bf x}_2 }
  G_0 ({\bf x}_1 - {\bf x}_1' ; L_1 ) G_0 ({\bf x}_2 - {\bf x}_2' ;L_2)
     \label{29}\end{eqnarray}
The integrals at coinciding end points can easily be performed,
and we find
\begin{eqnarray}
Z =
   \frac{2 \pi  M V^2}{(8 \pi )^3} (L_1 L_2)^{-3/2}
   \label{30}\end{eqnarray}

It is important to realize  that in Eq.~(\ref{15}),
 the limits of coinciding end points
${\bf x}_i' \rightarrow {\bf x}_i$ and
the inverse Laplace transformations
 do not commute unless a proper
renormalization scheme is chosen to eliminate the divergences
caused by the insertion of the composite operators $ |\psi (r)|^2$.
This can be seen for
a single polymer $P$. If we were to commuting the limit of coinciding end points with the Laplace
transform, we would obtain
\begin{eqnarray}
 &&\!\!\!\!\!\!\!\!\!\!\!\!\!\!\!\!\!
\!\!\!\!\!\!\!\!\!\!\!\!\int^{c + i\infty}_{c -i\infty} \frac{dz}{2\pi} {e^{zL}}
 \lim_{{\bf x}' \rightarrow {\bf x}} G_0 ({\bf x} -{\bf x}'; z)
\nonumber \\&    =
&\int^{c + i\infty}_{c-i\infty} \frac{dz}{2\pi i}e^{zL}G_0 ({\bf 0},z),
\label{31}\end{eqnarray}
where
\begin{eqnarray}
 G_0 ({\bf 0};z) = \langle |\psi ({\bf x})|^2\rangle .
\label{32}\end{eqnarray}
 This expectation value, however,
is linearly divergent:
\begin{eqnarray}
  \langle | \psi ({\bf x}_a) |^2 \rangle = \int
    \frac{d^3 k}{k^2 + m ^2} \rightarrow \infty
\label{33}\end{eqnarray}

\section{Calculation of Numerator in Second Moment}

Let us now turn to the numerator in Eq.~(\ref{1}):
\begin{eqnarray}
 N\equiv \int d^2{\bf x}_1 d^3 {\bf x}_2 \int ^{\infty}_{-\infty}
 dm ~m^2 G_m \left({\bf x}_1, {\bf x}_2 ; L_1, L_2\right).
\label{34}\end{eqnarray}
We set up  a functional integral for $N$ in terms of
the auxiliary probability
$
 G_{ \lambda =0} \left( \vec{\bf x}_1 ,\vec{\bf x}_2 ; \vec z    \right)
$
analogous to Eq.~(\ref{15}):
\begin{eqnarray}
   N  &= &\int d^3{\bf x}_1 d^3 r_2 \int^{\infty}_{-\infty} dm
 ~m^2 \lim_{{\bf x}'_1 \rightarrow {\bf x} \atop
    {\bf x}' \rightarrow {\bf x}_2}  \int^{c_\tau i\infty }_{c-i\infty}
   \frac{dz_i}{2\pi i}\frac{dz_2}{2\pi i} \nonumber \\
 & & e^{(z_1 L_1  + z_2 L_2)} \int^{\infty}_{-\infty}  d \lambda
      e^{-im \lambda }
 G_ \lambda  \left( \vec{\bf x}_1 ,\vec{\bf x}_2 ; \vec z    \right).
\label{35}\end{eqnarray}
The integration in $dm$ is easily performed after noting that
\begin{eqnarray}
\lefteqn{\!\!\!\!\!\!\!
  \int^{\infty}_{-\infty} dm ~m^2 e^{-im \lambda }
    G_ \lambda  \left( \vec{\bf x}_1 ,\vec{\bf x}_2 ; \vec z    \right)
}\nonumber \\
&& =
      - \int^{\infty}_{-\infty} dm \left(\frac{\partial^2}{\partial \lambda ^2}
 e^{-im \lambda }\right)
 G_ \lambda  \left( \vec{\bf x}_1 ,\vec{\bf x}_2 ; \vec z    \right).
\label{36}\end{eqnarray}
After a double  integration by parts in $  \lambda$,  and an integration
in $m$, we obtain
\begin{eqnarray}
  N & = &  \int d^3 {\bf x}_1  d^3 {\bf x}_2
    \lim_{{\bf x }_1' \rightarrow {\bf x}_1 \atop
     {\bf x}_2' \rightarrow {\bf x}_2} (-1)
    \int^{c + i\infty}_{c - i\infty} \frac{dz_1}{2\pi i}\frac{dz_2}{2\pi i}
    e^{(z_1 L_1 + z_2 L_2)} \nonumber \\
    & & \int^{\infty}_{-\infty} d \lambda ~  \delta ( \lambda )
	  \left[\frac{\partial^2}{\partial  \lambda ^2}
 G_ \lambda  \left( \vec{\bf x}_1 ,\vec{\bf x}_2 ; \vec z    \right)
   \right]
\label{37}\end{eqnarray}
Performing the now trivial integration in $d \lambda $
yields
\begin{eqnarray}
  N & = &  \int d^3 {\bf x}_1  d^3 {\bf x}_2
    \lim_{{\bf x }_1' \rightarrow {\bf x}_1 \atop
     {\bf x}_2' \rightarrow {\bf x}_2} (-1)
    \int^{c + i\infty}_{c - i\infty} \frac{dz_1}{2\pi i}\frac{dz_2}{2\pi i}
    e^{(z_1 L_1 + z_2 L_2)} \nonumber \\
   & &  ~~~~~~~~~~~~~~~~~\times
	  \left[\frac{\partial^2}{\partial  \lambda ^2}
 G_ \lambda  \left( \vec{\bf x}_1 ,\vec{\bf x}_2 ; \vec z    \right)
 \right]
   _{ \lambda =0}
\label{38}\end{eqnarray}
To compute     the term in brackets,
we
use again Eq.~(\ref{17}) and Eqs.~(\ref{18}) --(\ref{21}),
and find
\begin{eqnarray}
 N & = & \int d^3 {\bf x}_1 d^3 {\bf x}_2 \lim_{n_1 \rightarrow 0 \atop
    n_2 \rightarrow 0}
 \int^{c + i\infty}_{c - i \infty}  \frac{d z_1 }{2\pi i}
      \frac{dz_2}{2\pi i} e^{(z_1 L_1 + z_2 L_2)}
   \nonumber \\
 &&
    \times \int {\cal D}(\mbox{fields})~    \exp (-{\cal A}_0)
\vert\psi_1^{a_1} ({\bf x}_1)\vert^2
     \vert \psi_2^{a_2}  ({\bf x}_2 ) \vert^2
 \nonumber \\
 & &  \times \left[  \left(\int d^3 {\bf x}    \,{\bf A}_2 \cdot
 \Psi_2^*
{\bf \nablab} \Psi_2 \right)^2
 + \frac{1}{2} \int d^3 {\bf x} \, {\bf A}_2^2\,\vert \Psi_2 \vert^2
     \right].
\label{39a}\end{eqnarray}
In this equation we
have taken
the limits of coinciding endpoint
inside the
 Laplace integral over $z_1, z_2$.
This will be justified later
on the grounds that
the potentially dangerous Feynman diagrams containing
the insertions of operations like $|\Psi_i|^2$ vanish
in the limit $n_1, n_2 \rightarrow 0$.

\begin{figure}[tbhp]
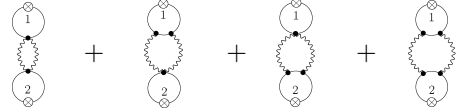

~\input 2u4s.tps~\\
\caption[Four diagrams contributing in Eq.~(\protect\ref{39}).
 The lines indicate correlation functions of $\Psi_i$-fields.
The crossed circles with label $i$  denote the insertion of
$|\psi_i^{a_i}({\bf x}_i)|^2$]{Four diagrams
contributing in Eq.~(\protect\ref{39}).
 The lines indicate correlation functions of $\Psi_i$-fields.
The crossed circles with label $i$  denote the insertion of
$|\Psi_i({\bf x}_i)|^2$.\label{4dia} }
\label{@FFF}\end{figure}

In order to calculate (\ref{39a}), we decompose the action  into a free
part
\begin{eqnarray}
{\cal A}^0_0 &\equiv  &
{\cal A}_{\rm CS}  \nonumber \\
&+& \int d^3 {\bf x} \left[ |{\bf D}^1 \Psi_1|^2 + | \nablab \Psi_2
  |^2 + \sum_{i=1}^22  |\Psi_i|^2 \right],
\label{18f}
\end{eqnarray}
 and  interacting parts
\begin{eqnarray}
{\cal A}^0_1
  \equiv \int d^3 {\bf x} \,\,{\bf j}_1{ ({\bf x}}) \cdot
   {\bf A}_1 ({\bf x})
\label{19i}\end{eqnarray}
 with     a
``current" of the first polymer field
\begin{eqnarray}
{\bf j}_1 ({\bf x}) \equiv  i \Psi_1^*({\bf x} ){ \nablab} \Psi_1({\bf x} ),
\label{20}
\end{eqnarray}
 and
\begin{eqnarray}
 {\cal A}_0^{2} \equiv \frac{1}{4} \int {d^3{\bf x}}~
    {\bf A}_1^2 |\Psi_1 ({\bf x})|^2 .
\label{21}\end{eqnarray}
Expanding the exponential
\begin{equation}
e^{ {\cal A}_0}=
e^{ {\cal A}_0^0+ {\cal A}_0^1+{\cal A}_0^2}=
e^{ {\cal A}_0}
\left[1\!-\! {\cal A}_0^1\!+\!\frac{({\cal A}_0^1)^2}2\!-\!{\cal A}_0^2\!+\!\dots~\right]
,
\label{@}\end{equation}
and keeping only the relevant terms,
the functional integral
(\ref{39a}) can be rewritten as a purely Gaussian expectation value
\begin{eqnarray}
 N & = & \kappa ^2 \int d^3 {\bf x}_1 d^3 {\bf x}_2 \lim_{n_1 \rightarrow 0 \atop
    n_2 \rightarrow 0}
 \int^{c + i\infty}_{c - i \infty}  \frac{d z_1 }{2\pi i}
      \frac{dz_2}{2\pi i} e^{(z_1 L_1 + z_2 L_2)}
   \nonumber \\
 &
    \times &\int {\cal D}(\mbox{fields})~    \exp (-{\cal A}_0^0)
\vert\psi_1^{a_1} ({\bf x}_1)\vert^2
     \vert \psi_2^{a_2}  ({\bf x}_2 ) \vert^2
 \nonumber \\
 &
\times& \left[  \left(\int d^3 {\bf x}    \,{\bf A}_1 \cdot
 \Psi_1^*
{\bf \nablab} \Psi_1 \right)^2
 \!\!+ \frac{1}{2} \int d^3 {\bf x} \, {\bf A}_1^2\,\vert \Psi_1 \vert^2
     \right]
   \nonumber \\
  &
\times& \left[  \left(\int d^3 {\bf x}    \,{\bf A}_2 \cdot
 \Psi_2^*
{\bf \nablab} \Psi_2 \right)^2
\!\! + \frac{1}{2} \int d^3 {\bf x} \, {\bf A}_2^2\,\vert \Psi_2 \vert^2
     \right]
\label{39}\end{eqnarray}
Only four diagrams shown in Fig.~(\ref{4dia})
contribute in
Eq.~(\ref{39}).%
Note that the initially asymmetric treatment of polymers
$P_1$ und
$P_2$
in the action (\ref{@5})
has led to a completely symmetric expression for the second moment.

\begin{figure}[tbhp]
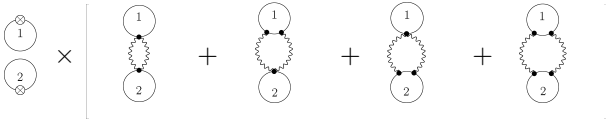


~~$\phantom{xx\Bigg(} $\input 2u4d.tps   $\phantom{\Bigg(}$ ~\\
\caption[Four diagrams contributing in Eq.~(\protect\ref{39}).
 The lines indicate correlation functions of $\Psi_i$-fields.
The crossed circles with label $i$  denote the insertion of
$|\Psi_i({\bf x}_i)|^2$]{Four diagrams contributing in Eq.~(\protect\ref{39}).
 The lines indicate correlation functions of $\Psi_i$-fields.
The crossed circles with label $i$  denote the insertion of
$|\Psi_i({\bf x}_i)|^2$. }
\label{4diad}\end{figure}

Only the first diagram in Fig.~\ref{4dia}
is divergent due to the divergence of the loop
formed by two vector correlation functions.
This infinity may be absorbed in the four-$\Psi$
interaction
accounting for the
excluded volume effect which we do not consider at the moment.
No divergence arises from the
insertiona of the composite fields
$\vert\Psi_i ({\bf x}_i)\vert^2$.
In this respect,
the disconnected diagrams shown in Fig.~\ref{4diad}
are
potentally dangerous.
But these vanish in the limit of zero replica indices
$n_1, n_2 \rightarrow 0$.

\section[Calculation of first Feynman diagrams in Fig.~\ref{@fff}]{Calculation of first Feynman diagrams in Fig.~\ref{@fff}}

From Eq.~(\ref{39}) one has to evaluate the following
integral
\begin{eqnarray}
~N_1\!\! & = &  \frac{ \kappa ^2}{4}\lim_{n_1 \rightarrow 0 \atop n_2 \rightarrow 0}
 \int^{c + i\infty }_{c - i \infty} \frac{d z_1}{2\pi i}
    \frac{dz_2}{2\pi i} e^{(z_1 L_1 + z_2 L_2)} \nonumber \\
 &&
           \int d^3
{x}_1  d^3 {x}_2
\int
     d^3 { x}_1' d^3 { x}_2'
  \label{40}\\
&& \bigg\langle \vert \psi_1^{a_1} ({\bf x}_1)\vert ^2
  \vert \psi_2^{a_2} ({\bf x}_2) \vert ^2
 \left(   \vert \Psi_1\vert^2 {\bf A}_1^2\right)_{{\sbf x}_1'}
\left(\vert \Psi_2\vert^2{{\bf A}_2^2}\right)_{{\sbf x}_2'}
 \bigg\rangle     . \nonumber
\end{eqnarray}
There is an ultraviolet-divergent  contribution which should be properly
regularized.
The system has, of course,
a
microscopic scale, which is
the size of the momomers.
This, however, is not the appropriate
short-distance scale to be uses here.
The model treats the polymers as random chains.
However, the momomers of
a polymer in the laboratory
are usually not freely movable,
so that polymers have a certain stiffness.
This gives rise to a certain
persistence length $\xi_0$
over which a polymer is stiff.
This length scale is increased to
$\xi>\xi_0$ by the excluded-volume effects.
This is the length scale which
should be used as a  proper physical short-distance
cutoff.
We may impose this cutoff by imagining the
model as being defined
on a simple cubic lattice of spacing $\xi$.
This would, of course, make analytical
calculations quite difficult.
Still, as we shall see,
it is possible to estimate the dependence
of the integral $N_1$ and the others
in the physically relevant limit in
which the lengths of the polymers are much larger than the persistence
length $\xi$.

An alternative and simpler regularization is based on
cutting off all ultraviolet-divergent
continuum integrals
at distances smaller than $\xi$.

After such a regularization, the calculation of
 $N_1$ is rather straightforward.
Replacing the expectation values by the
Wick contractions corresponding to the first diagram in
Fig.~\ref{4dia}, and performing the integrals
as shown in the Appendix, we obtain
\begin{eqnarray}
&&\!\!\!\!\!\!\!\!\!N_1 =\frac V{4\pi}\frac{M^2}{(8\pi)^6}(L_1L_2)^{-\frac
12}\nonumber \\&\times&
\int_0^1ds\left[(1-s)s\right]^{-\frac 3 2}
 \int d^3xe^{-\lfrac {M{\sbf x}^2}{2s(1-s)}}\label{ioneint}\\
&\times&
\int_0^1dt\left[(1-t)t\right]^{-\frac 3 2}
\int d^3ye^{-\lfrac {M{\sbf y}^2}{2t(1-t)}}\int
d^3x_1^{\prime\prime}\frac 1{\vert{\bf x}_1^{\prime\prime} \vert^4}  .
 \nonumber
\end{eqnarray}
%
The variables $\bf x$ and $\bf y$ have been rescaled with respect to
the original ones in order to extract the behavior of $N_1$ in $L_1$
and $L_2$. As a consequence, the lattices where $\bf x$ and $\bf
y$ are defined
have now spacings $ \xi/{\sqrt{L_1}}$ and $ \xi/{\sqrt{L_2}}$
respectively.

The ${\bf x},{\bf y}$ integrals may be explicitly computed in the
physical limit $L_1,L_2>>\xi$, in which the above spacings become
small.
Moreover, it is possible to approximate the integral in
${\bf x}_1^{\prime\prime} $ with an integral over a continuous
variable $\rho$ and a cutoff in the ultraviolet region:
\begin{eqnarray}
\int d^3x_1^{\prime\prime}\frac 1{\vert{\bf x}_1^{\prime\prime}
\vert^4}&\sim
&4\pi^2\int_\xi^\infty\frac{d\rho}{\rho^2}.\label{appsone}
 \label{@appr}\end{eqnarray}
After these approximations, we finally obtain
\begin{eqnarray}
N_1 = \frac{V   \pi ^{1/2}}{8} \frac{M^{-1}}{(8 \pi )^3}
    (L_1 L_2)^{-1/2}  \xi^{-1}.
\label{41}\end{eqnarray}

\section[Calculation of Second an Third Feynman Diagrams in
Fig.~\ref{@fff}]{Calculation of
Second
 an Third Feynman Diagrams in Fig.~\ref{@fff}}
Here we have to calculate
\begin{eqnarray}
&&
\!\!\!\!\!\!\!\!\!
\!\!\!\!\!\!\!\!\!\!\!\!\!
N_2
 =
 \kappa ^2  \lim_{n_1 \rightarrow 0 \atop n_2 \rightarrow 0}
     \int^{c + i \infty}_{c - i \infty} \frac{dz_1}{2\pi i}
\frac{dz_2}{2\pi i}
     e^{(z_1 L_1 + z_2 L_2)}
 \nonumber \\
 &&
         \!\!\!\!\!\!\!\!\!\!\!\!\!\times   \int d^3
{ x}_1  d^3 { x}_2
\int
     d^3 {x}_1'
d^3 {x}_1''
d^3 {x}_2'
\label{42}
\nonumber \\&&
 \!\!\!\!\!\!\!\!\!\!\!\!\!\times  \bigg\langle \vert \psi_1^{a_1}
({\bf x}_1)\vert^2
 \vert \psi_2^{a_2} ({\bf x}_2) \vert^2
 \left(
    \,{\bf A}_1 \cdot
 \Psi_1^*
{\bf \nablab} \Psi_1 \right)_{\sbf{x}_1'}
\nonumber \\
&& \!\!\!\!\!\!\!\!\!\!\!\!\!\times
 \left(
    \,{\bf A}_1 \cdot
 \Psi_1^*
{\bf \nablab} \Psi_1 \right)_{\sbf{x}_1''}
 \left( {\bf A}_2^2\,\vert \Psi_2 \vert^2\right)_{{\sbf x}_2'}
\bigg\rangle
\label{@56}\end{eqnarray}
The above amplitude has no ultraviolet divergence, so that no regularization is required.
The Wick contractions pictured in the second Feynman diagrams
of Fig.~\ref{4dia}
lead to the integral
\begin{eqnarray}
 N_2 = -2 V L_2^{-1/2} L_1^{-1} \frac{M}{(2\pi)^6} \int_{0}^{1}
 dt \int^{t}_{0} dt' C(t,t')
\label{43}
\label{@57}
\end{eqnarray}
 where $C(t,t')$ is a function independent of $L_1$ and $L_2$:
\begin{eqnarray}
  C(t,t') & = & \left[(1-t) t' (t-t') \right] ^{-3/2}
 \nonumber \\
& &
 \times \int
     d^3 { x} d^3 { y}  d^3 { z} e^{- M ({\sbf y} - {\sbf x })^2 /
     2(1-t)} \nonumber \\
 && \times \left( \nabla^ \nu_{\sbf y} e^{-M {\sbf y}^2 /2t'}\right) \left(
     \nabla^\mu_{\sbf x}  e^{-M {\sbf x}^2 /2 (t-t') }\right)\nonumber \\
 & & \times \frac{ \left[  \delta _{\mu  \nu } {\bf z} \cdot
     ({\bf z} + {\bf x}) - \left({ z} + { x}\right)_\mu
  { z}_ \nu \right]}
   {\vert{\bf z}\vert^3 \vert {\bf z} + {\bf x}\vert^3}.
\label{44}
\label{@58}
\end{eqnarray}
As in the previous section, the variables $\bf x, \bf y, \bf z$ have been
rescaled with respect to the original ones in order
to extract the behavior in $L_1$.

If the polymer lengths are much larger than the persistence length
one can ignore the fact that the monomers have a finite size and it is
possible to compute
$C(t,t')$ analytically, leading to 
\begin{eqnarray}
  N_2 & = & - \frac{V L_2^{-1/2}L_1^{-1} }{4^3 (2\pi)^6}
 M^{-1/2}   \sqrt{2} K,
\label{46}\end{eqnarray}
where $K$ is the constant
\begin{eqnarray}
K & \equiv & \frac{1}{6} B \left(\frac{3}{2},\frac{1}{2}\right) +
 \frac{1}{2} B\left(\frac{5}{2}, \frac{1}{2}\right)\nonumber \\& &
- B
 \left(\frac{7}{2}, \frac{1}{2}\right) + \frac{1}{3}
 B \left(\frac{9}{2}, \frac{1}{2}\right)   =\frac{19\pi}{384} \approx0.154
		 ,
\label{72}\end{eqnarray}
and $B(a,b) = { \Gamma (a)   \Gamma (b)}/{ \Gamma (a+b)}$ is the Beta function.
For large  $L_1 \rightarrow \infty $, this diagram gives a negligible
contribution with respect to $N_1$.

The third diagram in Fig.~\ref{4dia}
give the same as the second, except that $L_1$ and $L_2$ are interchanged.
\begin{equation}
N_3=N_2|_{L_1\leftrightarrow  L_2}.
\label{46b}\end{equation}

\section[Calculation of Fourth Feynman Diagram in
Fig.~\ref{@fff}]{Calculation of
Fourth Feynman Diagram in Fig.~\ref{@fff}}
Here we have the integral
\begin{eqnarray}
~~ N_4 &=& - 4  \kappa ^2 \frac{1}{2}
  \lim_{ n_1 \rightarrow  0 \atop n_2 \rightarrow 0}
 \int^{c + i \infty}_{c - i \infty}  \frac{dz_1}{2\pi i}
    \frac{dz_2}{2 \pi i} e^{(z_1 L_1 + z_2 L_2)}\nonumber \\
&&\!\!\!\!\!\!\!\!\!\!\!\!\!\!\!\!\!\times   \int d^3 {x}_1 d^3 {x}_2
\int
     d^3 { x}_1' d^3 { x}_2'
     d^3 { x}_1'' d^3 {x}_2''  \label{504} \\&&
\!\!\!\! \!\!\!\!\!\!\!\!\!\!\!\!\!\times  \bigg\langle \vert \psi_1^{a_1}
({\bf x}_1)\vert^2
 \vert \psi_2 ({\bf x}_2^{a_2}) \vert^2
\! \left({\bf A}_1\! \cdot \!
 \Psi_1^*
{\bf \nablab} \Psi_1 \right)_{\sbf{x}_1'}
\! \left(
    {\bf A}_1 \!\cdot\!
 \Psi_1^*
{\bf \nablab} \Psi_1 \right)_{\sbf{x}_1''}
\nonumber \\
 &&  ~~~~~~ \phantom{xxxxx}
 \times \left({\bf A}_2\! \cdot\!
 \Psi_2^*
{\bf \nablab} \Psi_2 \right)_{\sbf{x}_2'}
 \left( {\bf A}_2 \cdot
 \Psi_2^*
{\bf \nablab} \Psi_2 \right)_{\sbf{x}_2''}  \bigg\rangle .\nonumber
\label{50}\end{eqnarray}
which has no ultraviolet divergence.
After some effort we find
\begin{eqnarray}
 && \!\!\!\!\!\!\!\!\!\!\!\!\!N_4 = -\frac{1}{2\cdot 4^6}
\frac{M^3V}{(2\pi)^{11}} (L_1L_2)^{-1/2}
  \nonumber \\
  &&\times   \int^{1}_{0} ds
  \int^{s}_{0} ds'
  \int^{1}_{0} dt
  \int^{t}_{0}  dt'
   C (s, s', t, t'),
\label{51}\end{eqnarray}
where
\begin{eqnarray}
   && C(s, s'; t,t') =  \left[ (1-s)s' (s-s') \right] ^{-3/2}
			 \left[ (1-t) t'  (t-t')\right] ^{-3/2}
   \nonumber \\
&& \times \int \frac{d^3 p}{ ( 2\pi )^3 } \left[
\epsilon_{ \mu \lambda  \alpha}
   \frac{p^ \alpha }{{\bf p}^2 }\epsilon_{ \nu  \rho  \beta}
   \frac{p^ \beta }{{\bf p}^2 } + \epsilon_{\mu \rho  \alpha}
    \frac{p^ \alpha }{{\bf p}^2  } \epsilon_{ \nu  \lambda  \beta}
     \frac{p^ \beta }{{\bf p}^2  }\right] \nonumber \\
& &\times  \left[ \int d^3 { x}' d^3 { y}' e^{-i \sqrt{L_1} {\sbf p}
     ({\sbf x}' - {\sbf y}') }e ^{-M{\sbf x} '{}^2 / 2(1-s)} \right.\nonumber \\
&&\times  \left.\left(\nabla_{{\sbf y}'}^ \nu  e^{-M{\sbf y}'{}^2 /2t'}\right)
    \left(\nabla_{{\sbf x}'}^{\mu} e^{-M ({\sbf x}- {\sbf y})^2 /2(s-s')}\right)
    \right] \nonumber \\
 & &\times \left[ \int d^3 { u}' d^3 { v}'
    e^{-i \sqrt{L_2} {\sbf p} ({\sbf u}' -{\sbf v}' )}
 e^{-M {\sbf v}'{}^2 /2 (1-t)} \right.\nonumber \\
&& \times \left. \left( \nabla^
\rho_{{\sbf u}'} e^{-M{\bf u} '{}^2 / 2t'}\right)
    \left(\nabla^ \lambda_{{\sbf v}'}
e^{
-M ({\bf u}' -{\bf v}' )^2 /2(t-t')}
 \right)\right]
\label{52}\end{eqnarray}
and ${\bf x}', {\bf y}'$ are scaled variables. To take
into account the finite persitence length,
they should be defined on a lattice with spacing
$\xi/ \sqrt{L_1}$.
Similarly, ${\bf u}', {\bf v}'$ should be considered on a lattice
with spacing $\xi/ \sqrt{L_2}$.
Without
performing the
space integrations
$d^3 {\bf x}' d^3 {\bf y}' d^3 {\bf u}' d^3 {\bf v}'$,
the behavior of $N_4$ as a function of the polymer lengths can be
easily estimated in the following limits:

1. $L_1 \gg 1; L_1 \gg L_2$
\begin{eqnarray}
N_4 \propto L_1^{-1}
\label{53}\end{eqnarray}

2. $L_2 \gg 1; L_2 \gg L_1$
\begin{eqnarray}
 N_4  \propto L_2^{-1}
\label{54}\end{eqnarray}

3. $L_1, L_2 \gg 1,~~ {L_2}/{L_1} =  \alpha  = \mbox{finite} $
\begin{eqnarray}
  N_4 \propto L_1^{-3/2}
\label{55}\end{eqnarray}

Moreover, if the lengths of the polymers are considerably larger than
the
persistence length, the function $C(s,s',t,t')$ can be computed in a
closed form:
%
%
\begin{eqnarray}
 N_4 &  \approx & - \frac{1}{(2\pi)^5} \frac{1}{(2\pi)^{3/2}}
\frac{32}{ \sqrt{2} }
 (L_1L_2)^{-1/2} M^{-1/2}V
 \nonumber \\
& \times &\int^{1}_{0} ds \int^{1}_{0} dt
    (1-s) (1-t) (st)^{1/2} \nonumber \\
&\times &\left[ L_1 t (1-s) + L_2 (1-t)s\right] ^{-1/2} .
  \label{57}\end{eqnarray}
It  is simple to check that this expression has exactly
the above
behaviors.

 \section{Final Result}
Collecting all contributions we obtain the result for the
second topological moment:
\begin{eqnarray}
 \langle m^2\rangle = \frac{N_1 + N_2 + N_3 +  N_4}{Z} ,
\label{58}\end{eqnarray}
with $N_1,\,N_2,\,N_3,\,N_4,\,Z$
given by Eqs.~(\ref{30}),
(\ref{41}),
(\ref{46}),
(\ref{46b}),
and
(\ref{57}).

In all formulas, we have assumed that
 the volume $V$ of the system is much larger
than the size of the volume occupied by a single polymer,
i.e.,
$V \gg L_1^3$

To discuss the physical content of the result (\ref{58}),
we assume $P_2$ to be a long effective polymer
representing all polymers in a uniform solution.
We introduce the polymer concentration
$\rho $ as the average mass density  of the polymers per unit volume:
\begin{eqnarray}
\rho = \frac{M}{V}
\label{65}\end{eqnarray}
where $M$ is the total mass of the polymers
\begin{eqnarray}
 M = \sum_{i = 1}^{N_p} m_a \frac{L_k}{a}.
\label{66}\end{eqnarray}
Here
$m_a $ is the mass of a single monomer of length $a$,
$L_k  $ is the length of polymer $P_k$, and
$N_p  $ is the total number of polymers.
Thus ${L_k}/{a}$ is the  number of monomers in the
polymer $P_k$.
The polymer $P_1$ is singled out as any of the polymers $P_k$, say $P_{\bar k}$,
of length $L_1=L_{\bar k}$.
The
remaining ones are replaced by a long effective polymer  $P_2$
of length $L_2= \sigma_{k\neq \bar k} L_k$.
From the above relations we may also write
\begin{eqnarray}
  L_2 \approx \frac{a V \rho }{m_a}
\label{69}\end{eqnarray}
In this way, the length of the effective molecule
$P_2$ is expressed in terms of physical parameters,
the concentration of polymers, the monomer length,
and the mass and volume of the system.
Inserting (\ref{69}) into
(\ref{58}), with
 $N_1,\,N_2,\,N_3,\,N_4,\,Z$
given by Eqs.~(\ref{30}),
(\ref{41}),
(\ref{46}),
(\ref{46b}),
and
(\ref{57}).
and keeping only the
leading terms for $V \gg 1$,
we find for
 the second topological moment $\langle m^2\rangle$
the approximation
\begin{eqnarray}
 \langle m^2 \rangle \approx \frac{N_1 + N_2}{Z},
\label{70}\end{eqnarray}
and this has the approximate form
\begin{eqnarray}
\langle m^2 \rangle = \frac{a\rho }{m_a} \left[
   \frac{\xi^{-1} L_i}{16 \pi ^{1/2} M^2}-
     \frac{K  \sqrt{2}L_1^{1/2} }{(2\pi)^4 M^{3/2}} \right],
\label{71}\end{eqnarray}
with $K$ of (\ref{72}).
\section{Summary}
We have set up a topological field theory
to describe two fluctuating  polymers $P_1$ and $P_2$, and calculated
the second topological moment
for the linking
number $m$ between  $P_1$ and $P_2$.
The result is used to calculate the second moment for a single polymer
with respect to all others in a solution of many polymers.

In forthcoming work we shall
 calculate the effect of the excluded volume.

\section{Appendix}
In this appendix we present the computations of the amplitudes
$N_{1}, \dots, N_4$.
We shall need the
 following simple tensor
formulas involving  two completely antisymmetric
tensors $\varepsilon^{\mu \nu  \rho } $:
\begin{eqnarray}
\varepsilon_{\mu \nu  \rho } \varepsilon^{\mu \alpha  \beta }
 =  \delta _ \nu ^ \alpha   \delta _ \rho ^ \beta  -  \delta _ \nu ^ \beta
      \delta _ \rho ^ \alpha
\label{A-1}\end{eqnarray}
\begin{eqnarray}
  \varepsilon_{\mu \nu  \rho }\varepsilon^{\mu \nu  \beta }
  = 2  \delta _ \rho ^ \beta  .
\label{A-2}\end{eqnarray}
The Feynman diagrams shown in Fig.~\ref{4dia})
corresponds to a product of four correlation functions
$G_0$
of  Eq.~(\ref{33}), which have to be integrated over space
and Laplace transformed.
For the latter
we make use of the convolution
property
 of the integral over two
 Laplace
 transforms
 $\tilde f (z) $ and $\tilde g (z) $
of
the functions $f,g$:
\begin{eqnarray}
 \int^{c + i \infty}_{c - i \infty}
 \frac{dz}{2ni} e^{zL}
    \tilde f (z) \tilde g (z) = \int^{L}_{0} ds f(s)
    g (L-S)
\label{A-3}\end{eqnarray}
All spatial integrals are Gaussian of the form
\begin{eqnarray}
 \int d^3 x e^{-a {\bf x} + 2 b {\bf  x } \cdot {\bf y}} =
  (2n)^{3/2} a^{-3/2} e^{b^2 y^2/a} ,~~~a > 0.
\label{A_4}\end{eqnarray}
We are now ready to evaluate $N_1$ in Eq.~(\ref{40}).
Taking
 the limit of vanishing replica indices
we find with the help of Eqs.~(\ref{A-1})-(\ref{A-3}):
\begin{eqnarray}
 N_1 & = &\! \int d^3 x_1, d^3 x_2 \!\int^{L_1}_{0} ds \int^{L_2}_{0}
  \!\! dt\! \int d^3 x_1' d^3 x_2'
 \nonumber \\
 & \times&
G_0 ({\bf x}_1\! -\! {\bf x}_1' ; s) G_0 ({\bf x}_1' - {\bf x}_1 ; L_1 - S)
 \nonumber \\
 & \times&  G_0 ({\bf x}_2 -
      {\bf x}_2' ; t) G_0 ({\bf x}_2' -  {\bf x}_2 ; L_2 -t)
  \frac{l}{|{\bf x}_1' - {\bf x}_2'| s}.
\label{A_5}\end{eqnarray}
Performing the  changes of variables
\begin{eqnarray}
 s' = \frac{s}{L_1}~~~ t' = \frac{t}{L_2} ~~~{\bf x} = \frac{{\bf x}_1
    {\bf x}_1'}{ \sqrt{4} }~~~ {\bf y} = \frac{{\bf x}_2-{\bf x}_2'}
{( \sqrt{L_2} )}
\label{A_6}\end{eqnarray}
and setting ${\bf x}_1'' \equiv {\bf x}_1 ' - {\bf x}_2'$,
we easily derive
(\ref{ioneint}).

For small $\lfrac{\xi}{ \sqrt{L_1} }$ and $\lfrac{\xi}{ \sqrt{L_2} }$,
we d use the approximation (\ref{@appr}), the space
integrals can be done
the formula (\ref{A_4}).
After some calculation one finds the final result of Eq.~(\ref{54}).

The  amplitude $N_2$ calculated quite
similarly.
Contracting the fields in Eq.~(\ref{@56}), and keeping only the
contributions which do not vanish in the limit of zero replica indices,
we arrive at
\begin{eqnarray}
  N_2  & = & \int d^3 x_1 d^3 x_2 \int d^3 x_1' d^3 x_1'' d^3 x_2'
   \nonumber \\
   &\times &   \left[ \int_{0}^{L_1} ds \int^{S}_{0} ds' G_0 ({\bf x}_1'
- {\bf x}_1 ; L_1 - s) \right. \nonumber \\
&\times & \left. \nabla^{ \nu }_{x_1''} G_0 ({\bf x}_1 - {\bf x}_1'' ;
 s') \nabla^{\mu}_{x'_1} G_0 ({\bf x}_1'' - {\bf x}_1';
    s - s')\right]  \nonumber \\
&\times & D_{\mu \lambda }  ({\bf x}_1' - {\bf x}_2) D_ {\nu \lambda } ({\bf x}_1''
   - {\bf x}_2') \nonumber \\
&\times & \left[ \int^{L_2}_{0}dt G_0 ({\bf x}_2 - {\bf x}_2'; L_2 - t)
 G_0 ({\bf x}_2' - {\bf x}_2 ; t)\right].
\label{A_7}\end{eqnarray}
where $D_{\mu \nu }({\bf x},{\bf x}')$ are the correlation functions
(\ref{@7})--(\ref{@9})
of the vector potentials.
Setting
 $ {\bf x}_2 \equiv   \sqrt{L_2} {\bf u} + {\bf x}_2'$ and supposing
that $\lfrac{\xi}{ \sqrt{L_2} } $ is small, the integral
over ${\bf u}$ can be easily evaluated with the help
of the Gaussian integral (\ref{A_4}).
After the substitutions
$ {\bf x}_1'' =  \sqrt{L_1} {\bf y} + {\bf x}_1$ ~~
${\bf x}_1' =  \sqrt{L_1} ({\bf y}-{\bf x}) + {\bf x}_1$,~~
$ {\bf x}_2' =  \sqrt{L_1} ({\bf y} - {\bf x} - {\bf z}) + {\bf x}_1$
and a rescaling of the variables $s, s'$ by a factor $L_1^{-1}$, we
derive
Eq.~(\ref{@57}) with (\ref{@58}).

For small
 $\lfrac{\xi}{ \sqrt{L_1}}, \frac{3}{ \sqrt{L_2} } $,
the spatial integrals are easily evaluated leading to:
\begin{eqnarray}
 N_2& =& \frac{-  \sqrt{2} V L_2^{-1/2} L_1^{-1} M^{-1/2}}{(4n)^6}
    \nonumber \\
   &&   \int^{1}_{0} dt \int^{t}_{0} dt' t' (1-t)  \sqrt{\frac{t-t'}{1-t+t'}}
\label{A-8}\end{eqnarray}
 After the change of variable $t' \rightarrow
t'' = t - t'$,
 the double integral is reduced to a sum of integrals
the  type
\begin{eqnarray}
 c (n,m) = \int^{1}_{0} dt t^m \int^{t}_{0}  dt' t'{}^{n}
  \sqrt{\frac{t'}{ 1 - t'}},~~
m,n={\rm  integers}.\nonumber
\label{A-9}\end{eqnarray}
These can be simplified by replacing
$t^m$ by ${d}t^{m+1}/dt(m+1) $,
and doing the integrals by parts.
In this way, we end up with a linear combination of integrals of the form:
\begin{eqnarray}
   \int^{1}_{0} dt \frac{t^{\mu+ \frac{1}{2}}}{ \sqrt{1 - t} }
 =  B \left(\mu + \frac{3}{2}, \frac{1}{2}\right).
\label{A-10}\end{eqnarray}
The calculations of $N_3$ and $N_4$ are very similar, and may be
 omitted here.

\end{document}